\documentstyle[12pt,epsfig]{article}

\newcommand{\n}{\hspace*{-2.5mm}}
\newcommand{\arsinh}{\mathop{{\mbox{arsinh}}}\nolimits}
\newcommand{\arcosh}{\mathop{{\mbox{arcosh}}}\nolimits}
\newcommand{\li}{\mathop{{\mbox{Li}}_2}\nolimits}
\newcommand{\re}{\mathop{{\mbox{Re}}}\nolimits}
\newcommand{\im}{\mathop{{\mbox{Im}}}\nolimits}
\newcommand{\tr}{\mathop{{\mbox{tr}}}\nolimits}

\begin{document}
\title{\vskip-3cm{\baselineskip14pt
\centerline{\normalsize\hfill LMU--07/96}
\centerline{\normalsize\hfill MPI/PhT/96--055}
\centerline{\normalsize\hfill hep--ph/9607255}
\centerline{\normalsize\hfill July 1996}
}
\vskip1.5cm
Dispersion Relations in Loop Calculations\thanks{Lectures delivered at the
{\it XXXVI Cracow School of Theoretical Physics}, Zakopane, Poland,
June 1--10, 1996.}}
\author{{\sc Bernd A. Kniehl}\thanks{Permanent address:
Max-Planck-Institut f\"ur Physik (Werner-Heisenberg-Institut),
F\"ohringer Ring~6, 80805 Munich, Germany.}\\
{\normalsize Institut f\"ur Theoretische Physik,
Ludwig-Maximilians-Universit\"at,}\\
{\normalsize Theresienstra\ss e~37, 80333 Munich, Germany}}
\date{}
\maketitle
\begin{abstract}
These lecture notes give a pedagogical introduction to the use of dispersion
relations in loop calculations.
We first derive dispersion relations which allow us to recover the real part
of a physical amplitude from the knowledge of its absorptive part along the
branch cut.
In perturbative calculations, the latter may be constructed by means of
Cutkosky's rule, which is briefly discussed.
For illustration, we apply this procedure at one loop to the photon 
vacuum-polarization function induced by leptons as well as to the
$\gamma f\bar f$ vertex form factor generated by the exchange of a massive
vector boson between the two fermion legs.
We also show how the hadronic contribution to the photon vacuum polarization
may be extracted from the total cross section of hadron production in $e^+e^-$
annihilation measured as a function of energy.
Finally, we outline the application of dispersive techniques at the two-loop
level, considering as an example the bosonic decay width of a high-mass Higgs
boson.
\end{abstract}

\section{Introduction}

Dispersion relations (DR's) provide a powerful tool for calculating
higher-order radiative corrections.
To evaluate the matrix element, ${\cal T}_{fi}$, which describes the 
transition from some initial state, $|i\rangle$, to some final state, 
$|f\rangle$, via one or more loops, one can, in principle, adopt the following
two-step procedure.
In the first step, one constructs $\im{\cal T}_{fi}$ for arbitrary invariant
mass, $s=p_i^2$, by means of Cutkosky's rule \cite{cut}, which is a corollary
of $S$-matrix unitarity.
In the second step, appealing to analyticity, one derives $\re{\cal T}_{fi}$
by integrating $\im{\cal T}_{fi}$ over $s$ according to a suitable DR.

Dispersive techniques offer both technical and physical advantages.
Within perturbation theory, they allow us to reduce two-loop calculations
to standard one-loop problems plus phase-space and DR integrations,
which can sometimes be performed analytically even if massive particles are
involved \cite{sch,kni}.
This procedure can also be iterated to tackle three-loop problems \cite{kk}.
On the other hand, dispersive methods can often be applied where
perturbation theory is unreliable.
To this end, one exploits the fact that, by virtue of the optical theorem,
the imaginary parts of the loop amplitudes are related to total cross 
sections,
which may be determined experimentally as a function of $s$.
Perhaps, the best-known example of this kind in electroweak physics is
the estimation of the light-quark contributions to the photon vacuum
polarization---and thus to $\alpha(M_Z^2)$---based on experimental data of 
$\sigma(e^+e^-\to\mbox{hadrons})$ \cite{eid}.
This type of analysis may be extended both to higher orders in QED \cite{kkk}
and to a broader class of electroweak parameters \cite{pap}.

These lecture notes are organized as follows.
In Section~2, we derive DR's appropriate for physical amplitudes.
Cutkosky's rule is introduced in Section~3.
As an elementary application, we calculate, in Section~4, the leptonic
contribution to the photon vacuum polarization to lowest order in perturbation
theory, and relate its hadronic contribution to the total cross section of
$e^+e^-\to\mbox{hadrons}$.
In Section~5, we derive, from a Ward identity, a subtraction prescription for
general vacuum polarizations.
In Section~6, we evaluate, via a DR, the $\gamma f\bar f$ vertex form factor
generated by the exchange of a massive vector boson between the two fermion
legs.
In Section~7, we outline the application of DR's at the two-loop level,
considering the bosonic decay width of a high-mass Higgs boson.

\section{Dispersion relations}

In elementary particle physics, we often encounter form factors ({\it i.e.},
functions of $q^2$, where $q$ is some transferred four-momentum) which are
real-valued for $q^2$ below some threshold, $M^2$, and exhibit a branch cut for
$q^2>M^2$.
We shall discuss various examples of form factors in Sections~4--7.
In order to benefit from the powerful theorems available from the theory of
complex functions, it is necessary to allow for $q^2$ to be complex, although
this may contradict the na\"\i ve physical intuition.

\begin{figure}[ht]
\centerline{\epsfig{figure=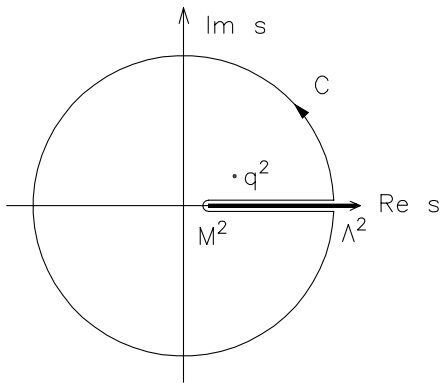,height=5cm}}
\caption{\label{fig1} Contour $\cal C$ of Eq.~(\ref{eqdis}) in
the complex-$s$ plane.}
\end{figure}

Let us then consider a complex-valued function, $F(s)$, of complex argument,
$s$, and assume that: (1) $F(s)$ is real for real $s<M^2$;
(2) $F(s)$ has a branch cut for real $s>M^2$;
(3) $F(s)$ is analytic for complex $s$ (except along the branch cut).
As usual, we fix the sign of the absorptive (imaginary) part of $F$ along the
branch cut by
\begin{equation}
F(s+i\epsilon)=\re F(s)+i\im F(s),
\end{equation}
where $\epsilon>0$ is infinitesimal.
By Schwartz' reflection principle, we then have
\begin{equation}
\label{eqsch}
F(s+i\epsilon)-F(s-i\epsilon)=2i\im F(s).
\end{equation}
Since $F$ is analytic at each point $q^2$ within contour $\cal C$ depicted in
Fig.~\ref{fig1}, we may apply Cauchy's theorem to find
\begin{eqnarray}
\label{eqdis}
F(q^2)&\n=\n&\frac{1}{2\pi i}\oint_{\cal C}ds\,\frac{F(s)}{s-q^2}\nonumber\\
&\n=\n&\frac{1}{2\pi i}\left(\int_{M^2}^{\Lambda^2}ds\,
\frac{F(s+i\epsilon)-F(s-i\epsilon)}{s-q^2}
+\oint_{|s|=\Lambda^2}ds\,\frac{F(s)}{s-q^2}\right)\nonumber\\
&\n=\n&\frac{1}{\pi}\int_{M^2}^{\Lambda^2}ds\,\frac{\im F(s)}{s-q^2-i\epsilon}
+\frac{1}{2\pi i}\oint_{|s|=\Lambda^2}ds\,\frac{F(s)}{s-q^2},
\end{eqnarray}
where we have employed Eq.~(\ref{eqsch}) in the last step.
Suppose that we only know $\im F$ along the branch cut and wish to evaluate
$F$ at some point $q^2$.
Then, Eq.~(\ref{eqdis}) is not useful for our purposes, since $F$ also appears
on the right-hand side, under the integral along the circle.
Thus, our aim is to somehow get rid of the latter integral.
If
\begin{equation}
\label{eqcon}
\lim_{\Lambda^2\to\infty}\oint_{|s|=\Lambda^2}ds\,\frac{F(s)}{s-q^2}=0,
\end{equation}
then we obtain the unsubtracted DR
\begin{equation}
\label{equns}
F(q^2)=\frac{1}{\pi}\int_{M^2}^{\infty}ds\,\frac{\im F(s)}{s-q^2-i\epsilon}.
\end{equation}
This means that $F$ can be reconstructed at any point $q^2$ from the knowledge
of its absorptive part along the branch cut.
In particular, the dispersive (real) part of $F$ may be evaluated from
\begin{equation}
\re F(q^2)=\frac{1}{\pi}\,{\cal P}\!
\int_{M^2}^{\infty}ds\,\frac{\im F(s)}{s-q^2},
\end{equation}
where $\cal P$ denotes the principal value.

Equation~(\ref{eqcon}) is not in general satisfied.
It is then useful to subtract from Eq.~(\ref{eqdis}) its value at some real
point $q_0^2<M^2$,
\begin{equation}
\label{eqsub}
F(q^2)=F(q_0^2)+\frac{q^2-q_0^2}{\pi}\int_{M^2}^{\Lambda^2}
\frac{ds}{s-q_0^2}\,\frac{\im F(s)}{s-q^2-i\epsilon}
+\frac{q^2-q_0^2}{2\pi i}\oint_{|s|=\Lambda^2}ds\,
\frac{F(s)}{(s-q_0^2)(s-q^2)}.
\end{equation}
If the last term in Eq.~(\ref{eqsub}) vanishes for $\Lambda^2\to\infty$, then
we have the once-subtracted DR
\begin{equation}
\label{eqonc}
F(q^2)=F(q_0^2)+\frac{q^2-q_0^2}{\pi}\int_{M^2}^{\infty}
\frac{ds}{s-q_0^2}\,\frac{\im F(s)}{s-q^2-i\epsilon}.
\end{equation}
Otherwise, further subtractions will be necessary.
For the use of DR's in connection with dimensional regularization, we refer to
Ref.~\cite{ner}.

\section{Cutkosky's rule}

In the previous section, we explained how to obtain the dispersive part of a 
form factor from its absorptive part.
Here, we outline a convenient method how to evaluate the absorptive part 
within perturbation theory.

Decomposing the scattering matrix as $S=1+iT$, where $T$ is the transition 
matrix, we obtain
\begin{equation}
\label{equni}
-i(T-T^\dagger)=T^\dagger T
\end{equation}
from the unitarity property $S^\dagger S=1$.
Since four-momentum is conserved in the transition from some initial state
$|i\rangle$ to some final state $|f\rangle$, we may always write
\begin{equation}
\label{eqt}
\langle f|T|i\rangle=(2\pi)^4\delta^{(4)}(P_f-P_i){\cal T}_{fi}.
\end{equation}
Consequently,
\begin{equation}
\langle f|T^\dagger|i\rangle=\langle i|T|f\rangle^*=
(2\pi)^4\delta^{(4)}(P_f-P_i){\cal T}_{if}^*.
\end{equation}
Inserting a complete set of intermediate states $|n\rangle$, we find
\begin{eqnarray}
\label{eqtt}
\langle f|T^\dagger T|i\rangle&\n=\n&\sum_n
\langle f|T^\dagger|n\rangle\langle n|T|i\rangle\nonumber\\
&\n=\n&(2\pi)^4\delta^{(4)}(P_f-P_i)\sum_n
(2\pi)^4\delta^{(4)}(P_n-P_i){\cal T}_{nf}^*{\cal T}_{ni}.
\end{eqnarray}
Using Eqs.~(\ref{eqt})--(\ref{eqtt}) in connection with Eq.~(\ref{equni}) and
peeling off the overall delta function, we obtain Cutkosky's rule,
\begin{equation}
\label{eqcut}
-i({\cal T}_{fi}-{\cal T}_{if}^*)=\sum_n(2\pi)^4\delta^{(4)}(P_n-P_i)
{\cal T}_{nf}^*{\cal T}_{ni},
\end{equation}
where it is understood that the sum runs over all kinematically allowed
intermediate states and includes phase-space integrations and spin summations.
Appealing to time-reversal invariance, we may identify the left-hand side of
Eq.~(\ref{eqcut}) with $2\im{\cal T}_{fi}$.
In summary, we may construct the absorptive part of a loop diagram according 
to the following recipe:
(1) cut the loop diagram in all kinematically possible ways into two pieces so
that one of them is connected to $|i\rangle$ and the other one to $|f\rangle$,
where cut lines correspond to real particles;
(2) stitch each pair of pieces together by summing over the spins of the real
particles and integrating over their phase space;
(3) sum over all cuts.

\section{Photon vacuum polarization}

One of the most straightforward applications of Cutkosky's rule and DR's is to
evaluate the one-loop photon vacuum polarization induced by fermions.
To avoid possible complications due to large nonperturbative QCD corrections,
we start by considering leptons.
For the sake of generality, we keep the electric charge, $Q$, and the number 
of colours, $N_c$, arbitrary.

\subsection{Leptonic contribution}

\begin{figure}[ht]
\centerline{\epsfig{figure=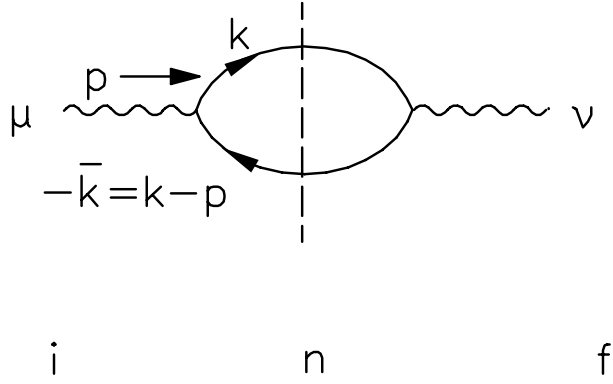,height=5cm}}
\centerline{\parbox{14cm}{\caption{\label{fig2}
Application of Cutkosky's rule to the lepton-induced photon vacuum
polarization.}}}
\end{figure}

Given the QED interaction Lagrangian, ${\cal L}_I=-eQ\bar\psi{\not\!\!A}\psi$,
we wish to compute ${\cal T}_{fi}$ depicted in Fig.~\ref{fig2}.
We start from the cut amplitudes,
\begin{eqnarray}
\label{eqtni}
i{\cal T}_{ni}&\n=\n&\bar u(k)(-ieQ)\gamma^\mu v(\bar k),\nonumber\\
i{\cal T}_{nf}&\n=\n&\bar u(k)(-ieQ)\gamma^\nu v(\bar k).
\end{eqnarray}
Summation over spins yields
\begin{eqnarray}
\sum_{spins}{\cal T}_{nf}^*{\cal T}_{ni}&\n=\n&
e^2Q^2\tr(\not\!\bar k-m)\gamma^\nu(\not\!k+m)\gamma^\mu\nonumber\\
&\n=\n&4N_ce^2Q^2
\left(k^\mu\bar k^\nu+\bar k^\mu k^\nu-\frac{s}{2}g^{\mu\nu}\right),
\end{eqnarray}
where $m$ is the fermion mass and $s=p^2=2(k\cdot\bar k+m^2)$.
Defining $d\tilde k=(d^3k/(2\pi)^32k^0)$, Eq.~(\ref{eqcut}) takes the from
\begin{eqnarray}
\label{eqimt}
2\im{\cal T}_{fi}&\n=\n&\int d\tilde kd\tilde{\bar k}\,(2\pi)^4
\delta^{(4)}(p-k-\bar k)\sum_{spins}{\cal T}_{nf}^*{\cal T}_{ni}\nonumber\\
&\n=\n&2N_c\alpha Q^2\sqrt{1-\frac{4m^2}{s}}T^{\mu\nu},
\end{eqnarray}
where $\alpha=(e^2/4\pi)$ is Sommerfeld's fine-structure constant and
\begin{equation}
T^{\mu\nu}=\int\frac{d\Omega}{4\pi}\,
\left(k^\mu\bar k^\nu+\bar k^\mu k^\nu-\frac{s}{2}g^{\mu\nu}\right).
\end{equation}
Upon integration, $p$ is the only four-momentum left, so that we can make the
ansatz $T^{\mu\nu}=Ap^\mu p^\nu+Bg^{\mu\nu}$.
In order to determine $A$ and $B$, we form the Lorentz scalars
\begin{eqnarray}
p_\mu p_\nu T^{\mu\nu}&\n=\n&s(sA+B)=0\nonumber\\
g_{\mu\nu}T^{\mu\nu}&\n=\n&sA+4B=-s-2m^2.
\end{eqnarray}
We so find $B=-A/s=-s-2m^2$, so that
$\im{\cal T}_{fi}=-(sg^{\mu\nu}-p^\mu p^\nu)\im\pi(s)$, where
\begin{equation}
\label{eqimp}
\im\pi(s)=\frac{N_c}{3}\alpha Q^2\left(1+\frac{2m^2}{s}\right)
\sqrt{1-\frac{4m^2}{s}}
\end{equation}
is the absorptive part of the photon vacuum polarization.
Notice that the dot product of ${\cal T}_{fi}$ with $p$ vanishes in compliance 
with electromagnetic gauge invariance.
Using the once-subtracted DR (\ref{eqonc}) with $q_0^2=0$, we obtain the
renormalized photon vacuum polarization as
\begin{eqnarray}
\label{eqpih}
\hat\pi(s)&\n=\n&\pi(s)-\pi(0)\nonumber\\
&\n=\n&\frac{s}{\pi}\int_{4m^2}^\infty\frac{ds^\prime}{s^\prime}\,
\frac{\im\pi(s^\prime)}{s^\prime-s-i\epsilon}\nonumber\\
&\n=\n&\frac{N_c}{3\pi}\alpha Q^2f\left(\frac{s+i\epsilon}{4m^2}\right),
\end{eqnarray}
where
\begin{equation}
\label{eqf}
f(r)=-\left(2+\frac{1}{r}\right)\sqrt{1-\frac{1}{r}}\arsinh\sqrt{-r}
+\frac{5}{3}+\frac{1}{r},
\end{equation}
appropriate for $r<0$.
This agrees with the well-known result found in dimensional regularization
\cite{hal}.
Representations of $f$ appropriate for $0<r<1$ and $r>1$ emerge from
Eq.~(\ref{eqf}) through analytic continuation.
Specifically, we have
\begin{equation}
\sqrt{1-\frac{1}{r}}\arsinh\sqrt{-r}
=\sqrt{\frac{1}{r}-1}\arcsin\sqrt{r}
=\sqrt{1-\frac{1}{r}}\left(\arcosh\sqrt{r}-i\frac{\pi}{2}\right).
\end{equation}
We verify that, for $r>1$, Eq.~(\ref{eqimp}) is recovered from
Eq.~(\ref{eqpih}).
The expansions of $f$ for $|r|\ll1$ and $r\gg1$ read
\begin{eqnarray}
\label{eqlim}
f(r)&\n=\n&\frac{4}{5}r+{\cal O}(r^2),\nonumber\\
\re f(r)&\n=\n&-\ln(4r)+\frac{5}{3}+{\cal O}\left(\frac{1}{r}\right),
\end{eqnarray}
respectively.
From Eq.~(\ref{eqlim}), we conclude that heavy fermions, with mass
$m\gg\sqrt{|s|}/2$, decouple from QED \cite{app}, while light fermions, with
$m\ll\sqrt{|s|}/2$, generate large logarithmic corrections.
The latter point creates a principal problem for the estimation of the
hadronic contribution to $\hat\pi$.
The evaluation of Eq.~(\ref{eqpih}) using the poorly known light-quark masses,
$m_q$, would suffer from large uncertainties proportional to $\delta m_q/m_q$.
In addition, there would be large nonperturbative QCD corrections in 
connection with the subtraction term $\pi(0)$.
In the next section, we discuss an elegant way to circumvent this problem.

\subsection{Hadronic contribution}

Let us consider the creation of a quark pair by $e^+e^-$ annihilation via a 
virtual photon, $e^-(l)e^+(\bar l)\to\gamma^*(p)\to q(k)\bar q(\bar k)$.
The corresponding $T$-matrix element reads
\begin{equation}
i{\cal T}=\bar v(\bar l)(-ieQ_e)\gamma^\mu u(l)\frac{-ig_{\mu\nu}}{s+i\epsilon}
i({\cal T}_{ni})^\nu,
\end{equation}
where $s=p^2$ and $i({\cal T}_{ni})^\mu$ is given in Eq.~(\ref{eqtni}).
Taking into account the $e^\pm$ spin average and the flux factor, we evaluate
the total cross section as
\begin{eqnarray}
\label{eqsig}
\sigma(s)&\n=\n&\frac{1}{4}\,\frac{1}{2s}\int d^3\tilde kd^3\tilde{\bar k}\,
(2\pi)^4\delta^{(4)}(l+\bar l-k-\bar k)\sum_{spins}|{\cal T}|^2\nonumber\\
&\n=\n&\frac{e^2Q_e^2}{8s^3}\,\tr\not l\gamma_\nu\not\bar l\gamma_\mu
\int d^3\tilde kd^3\tilde{\bar k}\,(2\pi)^4\delta^{(4)}(p-k-\bar k)
\sum_{spins}({\cal T}_{ni}^*)^\nu({\cal T}_{ni})^\mu\nonumber\\
&\n=\n&\frac{e^2Q_e^2}{8s^3}\,
4\left(l_\mu\bar l_\nu+\bar l_\mu l_\nu-\frac{s}{2}g_{\mu\nu}\right)
(-2)(sg^{\mu\nu}-p^\mu p^\nu)\im\pi(s)\nonumber\\
&\n=\n&\frac{e^2Q_e^2}{s}\,\im\pi(s),
\end{eqnarray}
where we have exploited Eq.~(\ref{eqimt}).
Substituting Eq.~(\ref{eqsig}) into Eq.~(\ref{eqpih}), we obtain
\begin{equation}
\label{eqhad}
\hat\pi(s)=\frac{s}{4\pi^2\alpha Q_e^2}\int_{4m^2}^\infty ds^\prime
\frac{\sigma(s^\prime)}{s^\prime-s-i\epsilon}.
\end{equation}
Equation~(\ref{eqhad}) allows us to estimate the hadronic contribution to 
$\hat\pi$ from the total cross section of hadron production by $e^+e^-$ 
annihilation measured as a function of $s$.
A recent analysis \cite{eid} has yielded
$-\hat\pi(M_Z^2)|_{hadrons}=0.0280\pm0.0007$.

\section{Subtraction prescription for general vacuum polarizations}

Vacuum polarizations have mass dimension two, so that the unsubtracted DR
(\ref{equns}) in general leads to ultraviolet divergences quadratic in the
cutoff $\Lambda$, which violate the Ward identities of the theory and are not
removed by renormalization.
It is therefore necessary to use a subtraction.
As we have seen in Section~3, in the case of QED, the na\"\i ve subtraction of
Eq.~(\ref{eqsub}) with $q_0^2=0$ leads to the correct physical result.
In the presence of unconserved currents, as in the Standard Model, the
situation is more complicated \cite{sch}.
In the following, we discuss a suitable subtraction prescription \cite{ks}.

Starting from the general interaction Lagrangian
${\cal L}_I=-B_\mu J^\mu+\mbox{h.c.}$, where $B^\mu$ is some vector field and
$J^\mu$ is the associated current, it is straightforward to derive the
vacuum-polarization tensor, which, by convention, differs from the $T$-matrix
element of $B^\mu(p)\to B^\nu(p)$ by a minus sign, as
\begin{equation}
\label{eqcor}
\Pi^{\mu\nu}(p)=-i\int d^4x\,e^{ip\cdot x}
\langle0|TJ^\mu(x)J^{\nu\dagger}(0)|0\rangle,
\end{equation}
where $T$ denotes the time-ordered product.
By Lorentz covariance, $\Pi^{\mu\nu}$ has the decomposition
\begin{equation}
\label{eqdec}
\Pi^{\mu\nu}(p)=\Pi(s)g^{\mu\nu}+\lambda(s)p^\mu p^\nu,
\end{equation}
where $s=p^2$.
Integrating by parts, we obtain from Eq.~(\ref{eqcor})
\begin{eqnarray}
\label{eqdel}
p_\mu\Pi^{\mu\nu}(p)&\n=\n&\int d^4x\,e^{ip\cdot x}
\langle0|T\partial_\mu J^\mu(x)J^{\nu\dagger}(0)|0\rangle\nonumber\\
&\n=\n&p^\nu\Delta(s),
\end{eqnarray}
where the last step follows from Lorentz covariance.
On the other hand, from Eq.~(\ref{eqdec}) we get
$p_\mu\Pi^{\mu\nu}(p)=p^\nu[\Pi(s)+s\lambda(s)]$.
Consequently, we have
\begin{equation}
\label{eqpdl}
\Pi(s)=\Delta(s)-s\lambda(s).
\end{equation}

By its definition (\ref{eqcor}), $\Pi$ has mass dimension two, so that its
evaluation from $\im\Pi$ via the unsubtracted DR (\ref{equns}) would be
quadratically divergent.
However, Eq.~(\ref{eqpdl}) relates $\Pi$ to quantities for which unsubtracted
DR's are only logarithmically divergent.
This is obvious for $\lambda$, which is dimensionless.
In the case of $\Delta$, this may be understood by observing that $J^\mu$ is
softly broken by mass terms and that one power of the external four-momentum
is extracted in Eq.~(\ref{eqdel}).
Writing unsubtracted DR's for $\Delta$ and $\lambda$, we find
\begin{eqnarray}
\label{eqks}
\Pi(s)&\n=\n&\frac{1}{\pi}\int_{M^2}^{\Lambda^2}
\frac{ds^\prime}{s^\prime-s-i\epsilon}
\left[\im\Delta(s^\prime)-s\im\lambda(s^\prime)\right]\nonumber\\
&\n=\n&\frac{1}{\pi}\int_{M^2}^{\Lambda^2}ds^\prime
\left[\frac{\im\Pi(s^\prime)}{s^\prime-s-i\epsilon}
+\im\lambda(s^\prime)\right],
\end{eqnarray}
where $M^2$ is the lowest threshold.
Detailed inspection at ${\cal O}(\alpha)$ and ${\cal O}(\alpha\alpha_s)$
\cite{ks} reveals that the logarithmic divergences of Eq.~(\ref{eqks}) exhibit
a similar structure as the poles in $\varepsilon=2-n/2$ found in
$n$-dimensional regularization, so that both methods lead to the same physical
results.

\section{Vertex correction}

\begin{figure}[ht]
\centerline{\epsfig{figure=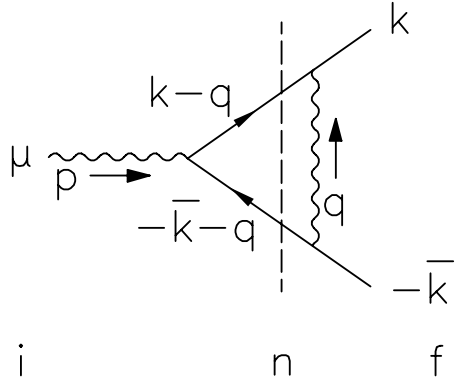,height=5cm}}
\centerline{\parbox{14cm}{\caption{\label{fig3}
Application of Cutkosky's rule to the $Bf\bar f$ vertex form factor induced by
the exchange of $B$ between the two fermion legs.}}}
\end{figure}

Cutkosky's rule (\ref{eqcut}) is also applicable if the initial and final 
states are different.
As an example, we now consider the interaction between a massless-fermion
current and a neutral vector boson, $B$, with mass $M$ characterized by the
Lagrangian ${\cal L}_I=-g\bar\psi{\not\!\!B}\psi$, where $g$ is the coupling 
constant, and evaluate the $Bf\bar f$ vertex form factor induced by the
exchange of $B$ between the two fermion legs as depicted in Fig.~\ref{fig3}.

The cut amplitudes read
\begin{eqnarray}
i{\cal T}_{ni}&\n=\n&\bar u(k-q)(-ig)\gamma^\mu v(\bar k+q),\nonumber\\
i{\cal T}_{nf}&\n=\n&-\bar u(k-q)(-ig)\gamma^\mu u(k)\,
\bar v(\bar k)(-ig)\gamma^\nu v(\bar k+q)
\frac{-ig_{\mu\nu}}{q^2-M^2+i\epsilon}.
\end{eqnarray}
Notice the extra minus sign of ${\cal T}_{nf}$.
The spin summation yields
\begin{equation}
\sum_{spins}{\cal T}_{nf}^*{\cal T}_{ni}
=\frac{g^3}{q^2-M^2}\bar u(k)\Gamma^\mu v(\bar k),
\end{equation}
where
$\Gamma^\mu=\gamma^\nu({\not\!k}-{\not\!q})\gamma^\mu({\not\!\bar k}+{\not\!q})
\gamma_\nu$.
Anticipating the multiplication with the $Bf\bar f$ tree-level amplitude, we
can bring $\Gamma^\mu$ into the form $\Gamma^\mu=N\gamma^\mu$, where
\begin{eqnarray}
N&\n=\n&-\frac{1}{4s}\tr\not\!k\Gamma^\mu\not\!\bar k\gamma_\mu\nonumber\\
&\n=\n&\frac{8}{s}k\cdot(\bar k+q)\bar k\cdot(k-q)
\end{eqnarray}
and $s=p^2$.
The integration over the phase space of the cut particles, with four-momenta
$k_+=\bar k+q$ and $k_-=k-q$, is most conveniently carried out in the
centre-of-mass frame using the variables $p=k_++k_-$ and $r=k_+-k_-$.
Then, we have $q^2=-s(1+z)/2$ and $N=s(1-z)^2/2$, where $z$ is the cosine
of the angle between $\bf r$ and $\bf k$.
Cutkosky's rule (\ref{eqcut}) now takes the form
\begin{eqnarray}
\im{\cal T}_{fi}&\n=\n&\frac{1}{2}\int d\tilde k_+d\tilde k_-\,(2\pi)^4
\delta^{(4)}(p-k_+-k_-)\sum_{spins}{\cal T}_{nf}^*{\cal T}_{ni}\nonumber\\
&\n=\n&-g\bar u(k)\gamma^\mu v(\bar k)\frac{g^2}{16\pi^2}\im F(s),
\end{eqnarray}
where
\begin{eqnarray}
\label{eqimf}
\im F(s)&\n=\n&\frac{\pi}{2}\int_{-1}^{1}dz\,\frac{(1-z)^2}{1+z+2/x}
\nonumber\\
&\n=\n&\pi\left[2\left(1+\frac{1}{x}\right)^2\ln(1+x)-3-\frac{2}{x}\right]
\end{eqnarray}
and $x=s/M^2$.
Finally, we obtain the renormalized vertex function through the 
once-subtracted DR (\ref{eqonc}) as
\begin{eqnarray}
\label{eqfs}
\hat F(s)&\n=\n&F(s)-F(0)\nonumber\\
&\n=\n&\frac{s}{\pi}\int_0^\infty\frac{ds^\prime}{s^\prime}\,
\frac{\im F(s^\prime)}{s^\prime-s-i\epsilon}\nonumber\\
&\n=\n&2\left(1+\frac{1}{x}\right)^2\left[\li(1+x)-\zeta(2)\right]
+\left(3+\frac{2}{x}\right)\ln(-x)-\frac{7}{2}-\frac{2}{x},
\end{eqnarray}
where $\li(z)=-\int_0^1dt\,\ln(1-zt)/t$ is the dilogarithm.
This agrees with the corresponding result found in dimensional regularization 
\cite{bak}.
Notice that, in Eq.~(\ref{eqfs}), $x=(s+i\epsilon)/M^2$ comes with an 
infinitesimal imaginary part.
The dispersive and absorptive parts of $F$ for $x>0$ are conveniently
separated by using the relations
\begin{eqnarray}
\ln(-x)&\n=\n&\ln x-i\pi,\nonumber\\
\li(1+x)-\zeta(2)&\n=\n&-\li(-x)-\ln(-x)\ln(1+x).
\end{eqnarray}
We so recover Eq.~(\ref{eqimf}), which serves as a useful check.

\section{Two-loop application}

\begin{figure}[ht]
\centerline{\epsfig{figure=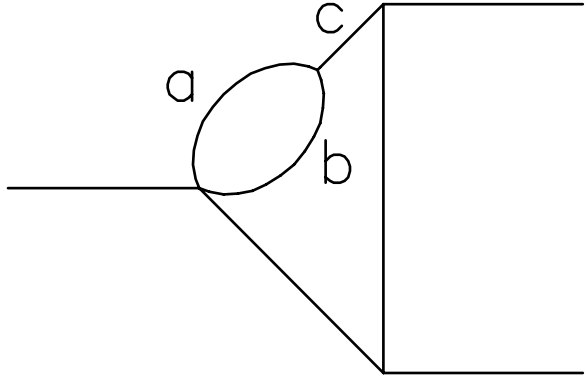,height=5cm}}
\centerline{\parbox{14cm}{\caption{\label{fig4}
Massive scalar two-loop three-point diagram, which contributes to the
Higgs-boson decay to a pair of intermediate bosons.
$a$, $b$, and $c$ denote the squared masses of the respective loop 
particles.}}}
\end{figure}

In this final section, we illustrate the usefulness of DR's beyond one loop,
considering the massive scalar two-loop three-point diagram depicted in 
Fig.~\ref{fig4}.
The idea is to write the one-loop two-point subdiagram contained in that
diagram as a once-subtracted DR and to interchange the loop and DR 
integrations, so as to reduce the two-loop problem at hand to a one-loop one
with a subsequent DR integration \cite{fri}.

In dimensional regularization, the scalar one-loop two-point integral is given 
by
\begin{eqnarray}
\label{eqb}
{\cal B}(s,a,b)&\n=\n&\left(\frac{\mu^2e^\gamma}{4\pi}\right)^\varepsilon
\int\frac{d^nq}{(2\pi)^n}\,\frac{1}{(q^2-a+i\epsilon)
\left[(q+p)^2-b+i\epsilon\right]}\nonumber\\
&\n=\n&\frac{i}{(4\pi)^2}e^{\varepsilon\gamma}\Gamma(\varepsilon)\int_0^1
\frac{dx}{X^\varepsilon}\nonumber\\
&\n=\n&\frac{i}{(4\pi)^2}e^{\varepsilon\gamma}\Gamma(1+\varepsilon)
\left[\frac{1}{\varepsilon}+f(s)+{\cal O}(\varepsilon)\right],
\end{eqnarray}
where $\gamma$ is the Euler-Mascheroni constant,
$\Gamma$ is the gamma function,
$n=4-2\varepsilon$ is the dimensionality of space time,
$\mu$ is the 't~Hooft mass scale introduced to keep the coupling constants
dimensionless,
$s=p^2$,
$X=\left[(1-x)a+xb-x(1-x)s-i\epsilon\right]/\mu^2$, and
$f(s)=-\int_0^1dx\,\ln X$.
The peculiar form of the prefactor in Eq.~(\ref{eqb}) is to suppress the
appearance of the familiar terms proportional to $\gamma-\ln(4\pi)$ in the
expressions.
The absorptive part of $f$ is
\begin{eqnarray}
\im f(s)&\n=\n&\pi\int_0^1dx\,\theta\Bigl(x(1-x)s-(1-x)a-xb\Bigr)\nonumber\\
&\n=\n&\pi\frac{\sqrt{\lambda(s,a,b)}}{s}
\theta\left(s-\left(\sqrt a+\sqrt b\right)^2\right),
\end{eqnarray}
where $\lambda(s,a,b)=s^2+a^2+b^2-2(sa+ab+bs)$ is the K\"all\'en function.
Using Eq.~(\ref{equns}), we then find
\begin{eqnarray}
{\cal B}(s,a,b)-{\cal B}(c,a,b)
&\n=\n&\frac{i}{(4\pi)^2}e^{\varepsilon\gamma}\Gamma(1+\varepsilon)
\left[f(s)-f(c)+{\cal O}(\varepsilon)\right]\nonumber\\
&\n=\n&\frac{i}{(4\pi)^2}e^{\varepsilon\gamma}\Gamma(1+\varepsilon)
\left[\frac{1}{\pi}\int_0^\infty d\sigma\,\im f(\sigma)\left(
\frac{1}{\sigma-s-i\epsilon}-\frac{1}{\sigma-c-i\epsilon}\right)\right.
\nonumber\\
&\n\n&{}+\left.\vphantom{\frac{1}{\pi}}{\cal O}(\varepsilon)\right].
\end{eqnarray}
Consequently, the subdiagram in Fig.~\ref{fig4} consisting of the bubble and
the adjacent propagator with four-momentum $q$ and squared mass $c$ may be
written as \cite{fri}
\begin{eqnarray}
\label{eqbub}
\frac{{\cal B}(q^2,a,b)}{q^2-c+i\epsilon}&\n=\n&
\frac{{\cal B}(c,a,b)}{q^2-c+i\epsilon}
-\frac{i}{(4\pi)^2}e^{\varepsilon\gamma}\Gamma(1+\varepsilon)
\int_{\left(\sqrt a+\sqrt b\right)^2}^\infty
\frac{d\sigma}{\sigma-c-i\epsilon}\,\frac{\sqrt{\lambda(\sigma,a,b)}}{\sigma}\,
\frac{1}{q^2-\sigma+i\epsilon}\nonumber\\
&\n\n&{}+{\cal O}(\varepsilon).
\end{eqnarray}
If Eq.~(\ref{eqbub}) is inserted in the expression for the one-loop seed
diagram in Fig.~\ref{fig4}, the first term turns into a product of two one-loop
diagrams, which contains all divergences.
In the second term, we may interchange the DR and loop integrations and are
left with a finite dispersion integral, which may be solved analytically.

\bigskip
\centerline{\bf ACKNOWLEDGEMENTS}
\smallskip\noindent
The author is grateful to the organizers of the {\it XXXVI Cracow School of 
Theoretical Physics} for the perfect organization and the great hospitality.

\end{document}